\documentclass
[preprint,aps,prc,showpacs,tightenlines,preprintnumbers,amsmath,amssymb]
{revtex4}

\usepackage{CJK}                     
\usepackage[dvips]{graphicx}
\usepackage{bm}                      
\usepackage{mathptmx}                
\usepackage{dcolumn}                 
\usepackage{array}
\newcommand{\br}{{\bf r}}

\begin{document}

\title{Tensor correlation, pairing interaction and deformation in Ne isotopes and Ne hypernuclei}

\author{A. Li$^{1}$\footnote{liang@riken.jp}, E. Hiyama$^{1}$\footnote{hiyama@riken.jp}, X.-R. Zhou$^{2}$\footnote{xrzhou@xmu.edu.cn}, H.Sagawa$^{1,3}$\footnote{sagawa@u-aizu.ac.jp}}

\affiliation{$^1$RIKEN Nishina Center, RIKEN, Wako 351-0198, Japan\\
$^2$ Department of Physics and Institute of Theoretical Physics
and Astrophysics, Xiamen University, Xiamen 361005, China\\
$^3$ Center for Mathematics and Physics, University of Aizu,
Aizu-Wakamatsu, Fukushima 965-8560, Japan}

\date{\today}

\begin{abstract}

We study tensor and pairing effects on the quadruple deformation of
neon isotopes based on a deformed Skyrme-Hartree-Fock model with BCS
approximation for the pairing channel. We extend the
Skyrme-Hartree-Fock formalism for the description of single- and
double-lambda hypernuclei adopting two different hyperon-nucleon
interactions. It is found that the interplay of pairing and tensor
interactions is crucial to derive the deformations in several neon
isotopes. Especially, the shapes of $^{26,30}$Ne are studied in
details in comparisons with experimentally observed shapes.
Furthermore the deformations of the hypernuclei are compared with
the corresponding neon isotopic cores in the presence of tensor
force. We find the same shapes with somewhat smaller deformations
for single $\Lambda$-hypernuclei compared with their core
deformations. It is also pointed out that the latest version of
hyperon interaction, the ESC08b model, having a deeper $\Lambda$
potential makes smaller deformations for hypernuclei than those of
another NSC97f model.

\end{abstract}


\maketitle
\section{Introduction}

Recently, several neutron-rich neon isotopes are confirmed to be
deformed by the observation of the low excitation energies and
$B(E2)$ values of the first excited $2^+$
states~\cite{Ram01,Ril03,Yan03,Gei05,Iwa05,Gib07}, including
$^{30}$Ne with a $N = 20$ magic number. These empirical evidences
become a strong motivation to make extensive experimental and
theoretical studies on the neon
isotope~\cite{Sag04,Nak09,Ura11,Tak12,Min12,Sum12}. For example,
experimentally, Takechi {\it et al.}~\cite{Tak12} succeeded in
measuring the interaction cross section $\sigma_\mathrm{I}$ of
$^{20-32}$Ne isotopes and as a result, they have found that starting
mass number 25, the $\sigma_\mathrm{I}$ data exceed the systematic
mass-number dependence of $\sigma_\mathrm{I}$ for stable nuclei.
These observations indicate possible large nuclear deformations for
these nuclei. Particularly, the much enhanced $\sigma_\mathrm{I}$
data suggest that there are deformed halos for $^{29}$Ne and
$^{31}$Ne. The halo structure of $^{31}$Ne has been clearly observed
also by the measurement of Coulomb breakup cross section by Nakamura
et al.~\cite{Nak09}. Urata {\it et al.}~\cite{Ura11} then studied in
details this $^{31}$Ne halo structure in the framework of deformed
$^{30}$Ne + n two-body model. Moreover, two of the author in the
present work, H. S. and X. Z.~\cite{Sag04} studied the neon isotope
within the deformed Skyrme-Hartree-Fock (SHF) model with two Skyrme
interactions SGII and SIII, and investigated the neutron number
dependence of deformation properties along the chain of neon
isotopes. Reaction calculations of $^{20-32}$Ne isotopes by the
antisymmetrized molecular dynamics (AMD) have also been extensively
done~\cite{Min12,Sum12}, and they suggested large deformations for
most of the neutron-rich isotopes, consistent with the indication of
recent experiments.

However, some of the experimentally-determined shapes of isotopic
nuclei are not reproduced theoretically by many mean-field models.
To take $^{30}$Ne as an example, the observed low excitation energy
of $2^{+}_1$ state indicates a large deformation for this nuclei,
but both the Hartree-Fock-Bogoliubov (HFB) theory~\cite{Hil07} and
the relativistic HFB (RHFB) theory~\cite{Ebr11} give a spherical
result, as well as the above-mentioned SHF model~\cite{Sag04}. Since
the nuclear shape is closely related to its shell structure, a
tensor-force-driven deformation has been proposed
recently~\cite{Ots08,Uts09,Uts12} in the occurrence of large oblate
deformation in $^{42}$Si. Because the monopole interaction of tensor
force~\cite{Ots01,Ots05} will result in a smaller $1s1/2$-$0d5/2$
proton gap when more neutrons occupy $0f7/2$ state, the nucleus
favors energetically to be in an oblately-deformed state. This
indicates the crucial role of the tensor force on the deformation of
neutron-rich nuclei, even $N = 28$ is a magic number and supposed to
favor a spherical shape.

Many Skyrme parameter sets which have been widely used do not
include the tensor contribution, although it was suggested more than
50 years ago by Skyrme~\cite{Sky}. Only recently, a Skyrme
interaction which includes the tensor contribution was
proposed~\cite{Bro06}, and Col$\grave{o}$ {\it et al.}~\cite{Col07}
and Brink $\&$ Stancu~\cite{Bri07} have added Skyrme-type tensor
force on top of the existing standard parameterizations SLy5 and
SIII, respectively. After that, Lesinski {\it et al.}~\cite{Les07}
built a set of 36 effective interactions with a systematical
adjusting of the tensor coupling strengths. Each of these
parameterizations has been fitted to give a reasonable description
for ground states of finite nuclei such as binding energies and
radii. The inclusion of tensor terms in the calculations achieved
considerable success in explaining various nuclear structure
problems~\cite{Col07,Bri07,Les07,Gra07,Zou08,Zal08,Tar08,Bai09,Bai09r,Cao09,Tor10,Bai10,Bai11,Cao11,Bai11a}.
The present work is devoted to a systematic study of the tensor
effect on the deformation of neon isotopes, and theoretical results
will be confronted directly with recent experiments.

In addition, lately, the pairing interaction between nucleons has
been extended to be isospin-dependent~\cite{Mar07,Mar08}, and global
SHF calculations~\cite{Ber12} with an isospin-dependent contact
pairing interaction are shown to have a better agreement with
empirical data than the results without isospin-dependence. This may
be a hint that the pairing strength without the isospin dependence,
incorporated in the previous SHF study, should be reexamined taking
care of the isospin-dependence nature. This is because some
neutron-rich nuclei far from the line of stability have large
isospins, and the pairing strength might be largely reduced. It is
also known that the evolution of the deformation is rather
model-dependent especially on the role of the pairing
interaction~\cite{Hil07,Ebr11,Min12,Sum12}. Therefore as a first
step it is of great interest to investigate how the change of
pairing strength will influence the shape of nuclei. We will show in
this study that the pairing interaction has a decisive effect on the
nuclear deformation of several neutron-rich nuclei.

In hypernuclear physics, it is one of the interesting subjects to
study the modification of nuclear structure when a hyperon such as
$\Lambda$ particle is added to a nucleus. Theoretically, Motoba {\it
et al.}~\cite{Motoba} pointed out that sizes of $p$-shell $\Lambda$
hypernuclei are shrunk by the addition of a $\Lambda$ particle in
comparison with those of the core nuclei (It is called a gluelike
role of the $\Lambda$ particle). They suggested that the sizable
shrinkage effect was seen in the value of $B(E2)$. Afterwards, the
shrinkage effect on $B(E2)$ values was studied in the case of 5/2$^+
\rightarrow 1/2^+$ transition in $^7_{\Lambda}$Li~\cite{Hiy99}.
Then, the KEK-E419 experiment was successfully performed to measure
this $B(E2)$ and confirmed the shrinkage effect of $\Lambda$
particle on the nuclear size for the first time~\cite{Tan01}. Also,
some authors have studied the change of structure in heavier
hypernuclei by addition of a $\Lambda$
particle~\cite{Zho07,Win08,Lu11,Sch10,Isa11}. Within the framework
of mean-field models \cite{Zho07,Win08,Lu11}, they suggested the
changes of deformation between core nuclei in $^{12}$C and
$^{28,30,32}$Si, and the corresponding hypernuclei by the addition
of a $\Lambda$ particle. Furthermore it was pointed out from the
study of $^9_{\Lambda}$Be and $^{13}_{\Lambda}$C by the
AMD~\cite{Isa11} that if a nucleus occupies a $1p$ shell orbital, a
hyperon may enhance the nuclear deformation.

It should be noted that in hypernuclear physics there have been a
lot of ambiguities in proposed hyperon-nucleon ($YN$) interactions
due to the limitation of $YN$ scattering data. For a decade, by
combining the $\gamma$-ray experimental data and theoretical
calculations such as the shell model~\cite{mill} and the clustering
approach~\cite{Hiy00,Hiy06}, they succeeded in extracting
information on its spin-dependent parts of $\Lambda N$ interaction.
As a result, the most updated $YN$ interaction, Nijmegen soft core
potential such as the ESC08b model~\cite{Rij10} is proposed. In the
present work, we use this ESC08b potential to study the structure
changes of the Ne isotopes. In addition, we use the NSC97f
potential~\cite{Sto99} which reproduces the observed binding
energies of light $\Lambda$ hypernuclei. The main aim of this paper
is to study the importance of tensor and pairing effects on the
deformation properties of neon isotopes and the corresponding
hypernuclei.

The paper is organized as follows. In Sec. II, we outline the
necessary formalism. The numerical results and discussions are given
in Sec. III. Finally, Sec. IV contains the main conclusions and
future perspectives of this work.
\section{Formalism}

Our study is based on the deformed SHF method of Ref.~\cite{Vau73},
accompanied by a density-dependent contact pairing using the BCS
approximation. An extended model was proposed for the description of
hypernuclei in Refs.~\cite{Zho07,Cug00} by including an effective
microscopic lambda-nucleon interaction derived from
Brueckner-Hartree-Fock (BHF) calculations of isospin-asymmetric
nuclear matter using realistic Nijmegen YN potential~\cite{Schu}. In
the following we outline some necessary formalism.

The total binding energy of a nucleus can be obtained
self-consistently from the energy functional~\cite{Les07,Rei95}:
\begin{eqnarray}
\label{eq:eN} {\cal E}_N = \mathcal{E}_{\text{kin}}
  + {\cal E}_{\text{Sk}}
  + {\cal E}_{\text{pair}}
  + {\cal E}_{\text{Coul}}
  + {\cal E}_{\text{corr}}
\end{eqnarray}
where ${\cal E}_{\text{kin}}$ is the kinetic energy functional,
${\cal E}_{\text{Sk}}$ is the Skyrme functional, ${\cal
E}_{\text{pair}}$ is the pairing energy functional, ${\cal
E}_{\text{Coul}}$ is the Coulomb energy functional and ${\cal
E}_{\text{corr}}$ is the center-of-mass correction. More physics
details can be found in Refs.~\cite{Les07,Rei95} and references
therein, and we will only address some for ${\cal
E}_{\text{Skyrme}}$ and ${\cal E}_{\text{pair}}$, which are relevant
for the present study.

To generate the Skyrme energy functional ${\cal E}_{\text{Sk}}$, an
effective zero-range two-body tensor force~\cite{Sky} is included in
recent Skyrme parameterizations~\cite{Bro06,Col07,Bri07,Les07} as
follows:
\begin{eqnarray}
v_T(\br)& = & \frac{T}{2}\{[({\sigma}_1\cdot\textbf{k}^{'})({\sigma}_2\cdot\textbf{k}^{'})-\frac{1}{3}({\sigma}_1\cdot{\sigma}_2)\textbf{k}^{'2}]\delta(\br_1-\br_2) \nonumber \\
&   &  \quad \quad + \delta(\br_1-\br_2)
[({\sigma}_1\cdot\textbf{k})({\sigma}_2\cdot\textbf{k})-\frac{1}{3}({\sigma}_1\cdot{\sigma}_2)\textbf{k}^2]
\} \nonumber \\
&   & + U \{({\sigma}_1\cdot\textbf{k}^{'})\delta(\br_1-\br_2)
({\sigma}_1\cdot\textbf{k}) \nonumber \\
&   & \quad \quad -\frac{1}{3}({\sigma}_1\cdot{\sigma}_2)
[\textbf{k}^{'}\cdot\delta(\br_1-\br_2)]\textbf{k}\}
\end{eqnarray}
where $T$ and $U$ are the coupling constants denoting the strength
of the triplet-even and triplet-odd tensor interaction,
respectively. The operator $\textbf{k}\equiv (\nabla_1-\nabla_2)/2i$
acts on the right, and $\textbf{k}^{'}\equiv
-(\nabla_1-\nabla_2)/2i$ acts on the left. We will select several
effective interactions with tensor terms included from
Ref.~\cite{Les07} in the present study, to compare not only the
cases with or without tensor terms, but also the cases with
different tensor strengths.

In the Skyrme energy functional ${\cal E}_{\text{Sk}} = \int \! {\rm
d}^3 r \; {\cal H}^{\text{Sk}} (\br)$, the tensor part contributes
to the energy density ${\cal H}^{\text{Sk}}$ in a combined way with
the exchange term of central part as
\begin{eqnarray}
{\cal H}_T^{\text{Sk}} &=& \frac{1}{2}\alpha(J_n^2+J_p^2) + \beta\vec{J}_n\cdot\vec{J}_p; \\
\alpha     &=& \alpha_C+\alpha_T;  \quad \beta = \beta_C+\beta_T  \label{eq:t} \\
\alpha_C   &=& \frac{1}{8}(t_1-t_2) - \frac{1}{8}(t_1x_1+t_2x_2); \nonumber \\
\beta_C    &=& -\frac{1}{8}(t_1x_1+t_2x_2) \label{eq:t1} \\
\alpha_T   &=& \frac{5}{12}U;  \quad \beta_T = \frac{5}{24}(T+U).
\label{eq:t2}
\end{eqnarray}
indicated by a subscript of $T$ or $C$, respectively. $\alpha$
represents the strength of like-particle coupling between
neutron-neutron or proton-proton, and $\beta$ is that of the
neutron-proton coupling. The coupling strengths of various parameter
sets used in this study are collected in Tab.~1.

\begin{table}[t!]
\begin{center}
\caption{Coupling strengths (in MeV) of various parameter sets used
in the work.} \vspace{5pt}
\begin{tabular}{cccccccccccc}
\hline\hline
&~~T14~~&~~T24~~&~~T34~~&~~T44~~&~~T54~~&~~T61~~&~~T62~~&~~T63~~&~~T64~~&T65~~&~~T66~~
\\ \hline
$\alpha$ & 120 & 120 & 120 & 120 & 120 & -60& 0 & 60& 120 & 180& 240
\\ \hline
$\beta$ & -60 & 0 & 60 & 120 & 180 &240 &240 &240 & 240 & 240 & 240
\\ \hline
$\alpha_T$ & 38.5 & 24.7 & 12.8 & 8.97 & -3.48 & -200 & -131
&-80.5&-0.246 & 49.9 & 113
\\ \hline
$\beta_T$ & -15.2 & 19.4 & 57.8 & 113 & 150 &178 & 196 & 177 & 218 &
196 & 204
\\ \hline
$U$ & 92.5 & 59.2 & 30.8 & 21.5 & -8.36 & -480 & -314 & -193 &
-0.591 & 120 & 271
\\ \hline
$T$ & -165 & 33.7 & 247  & 521 & 727  &1044& 1256 &1335 & 1047 &823&
709
\\ \hline\hline
\end{tabular}\label{tensor}
\end{center}
\end{table}

The pairing energy functional ${\cal E}_{\text{pair}}$ is introduced
as:
\begin{equation}
\label{eq:PairFunc} {\cal E}_{\text{pair}} = \frac{1}{4}
\sum_{q\in\{p,n\}}
  \int \! {\rm d}^3r \; \chi_q^* (\br) \; \chi_q (\br) \;
  G(\br)
\end{equation}
where $\chi(\br)$ is the local pairing density matrix (addressed
later), and $G(\br)$ denotes the pairing strength function. We
choose $G(\br) = V_{\text{pair}}
\left(1-\frac{\rho(\br)}{\rho_0}\right)$ with $\rho_0 = 0.16\;\rm
fm^{-3}$, which corresponds to a density-dependent delta force for
the pairing interaction. Based on the empirical neutron pairing gaps
extracted by using the three-point mass difference
formula~\cite{Sat98} and the experimental binding energies of
Ref.~\cite{Aud03}, we choose $V_{\text{pair}}$ = 900 MeV, which can
reproduce reasonably the gap data for the whole isotopic chain. This
value is referred in the following as the full pairing cases
(labeled as $V_{\text{pair}}^{\rm full}$).

The local part of the pair density matrix $\chi(\br)$ is written as
\begin{eqnarray}
\label{eq:Pair} \chi_q (\br) = - 2 \sum_{k \in \Omega_q \atop k
> 0} f_k \, u_k \, v_k \,
      | \phi_k (\textbf{r}) |^2
\end{eqnarray}
with $q\in\{p,n\}$, and $\Omega_q$ is the configuration space
adapted. The $\phi_k$ are the singe particle (s.p.) wave functions
and $v_k$, \mbox{$u_k = \sqrt{1-v_k^2}$} are the pairing amplitudes.
The cutoff factors $f_k$ are included to prevent the unrealistic
pairing of highly excited states based on the employed pairing
energy functional of Eq.~(\ref{eq:PairFunc}), and more details are
referred to Ref.~\cite{Ben00}.

From the nuclear energy functional $\mathcal{E}_N$ of
Eq.~(\ref{eq:eN}), one can obtain the mean-field HF + BCS equations
for nucleons by standard functional derivative techniques, and they
are applied to calculate the ground-state properties of nuclei. For
the calculations of hypernuclei, the contribution due to the
presence of hyperons should be included accordingly:
\begin{eqnarray}
{\cal E} &=& \mathcal{E}_N + \mathcal{E}_{\Lambda};\label{eq:eL} \\
\mathcal{E}_{\Lambda} &=& \frac{\tau_{\Lambda}}{2m_{\Lambda}}
+(\frac{m_{\Lambda}}{m^*_{\Lambda}}-1)\frac{\tau_{\Lambda}-C\rho_{\Lambda}^{5/3}}{2m_{\Lambda}}
+\mathcal{E}_{N\Lambda};\\
\mathcal{E}_{N\Lambda} &=&
(\rho_N+\rho_{\Lambda})\frac{B}{A}(\rho_N,\rho_{\Lambda})-\rho_N\frac{B}{A}(\rho_N,0)-\frac{C\rho_{\Lambda}^{5/3}}{2m_{\Lambda}} \nonumber \\
&   &  \nonumber \\
\end{eqnarray}
where the energy density functional $\mathcal{E}_{N\Lambda}$ is
obtained from a fit to the binding energy per baryon,
$B/A(\rho_N,\rho_{\Lambda})$ of asymmetric hypermatter, generated by
BHF calculations~\cite{Schu}. The adequate $\Lambda$ effective mass
$m^*_{\Lambda}$ is computed from the BHF single-$\Lambda$ potential
obtained in the same calculations. Then the HF equations for a
hypernuclei system can be routinely obtained from the total energy
functional of Eq.~(\ref{eq:eL}), as detailed in
Ref.~\cite{Zho07,Cug00}.

In this work two kinds of Nijmegen soft-core hyperon potentials are
used in BHF calculations for the derivation of effective microscopic
lambda-nucleon interactions: one is the NSC97f model~\cite{Sto99},
another is the recently developed ESC08b model~\cite{Rij10}. The
ESC08b potential contains no hyperon-hyperon (YY) components,
whereas the NSC97f potential comprises the extension to the YY
sector based on SU(3) symmetry. That is, the pairing interaction
between $\Lambda$ hyperons is considered only in the study with the
NSC97f potential. The NSC97f potential in the past has been
preferred for the study of $\Lambda$ hypernuclei since it reproduces
rather well the available experimental binding data of $\Lambda$
hypernuclei~\cite{Yamamoto}. The ESC08b potential, on the other
hand, has been demonstrated to have a slightly more attractive
$\Lambda$ potential in nuclear matter, with $-40$ MeV~\cite{Sch11}
for ESC08b and $-36$ MeV~\cite{Vid01} for NSC97f, respectively. One
can thus expect that it is harder to evolve deformed $\Lambda$
hypernuclei with the ESC08b, which is confirmed later by the
comparison between the NSC97f and ESC08b results.

\section{Results}

\begin{figure}[t!]
\centering
\includegraphics[width=13.5cm]{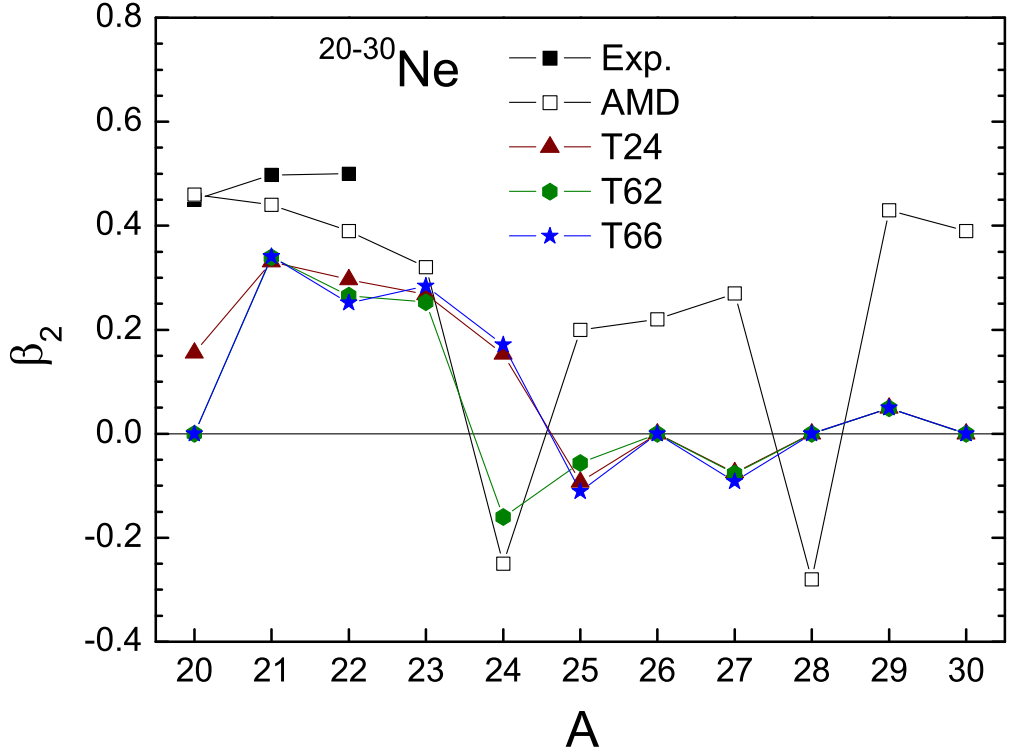}
\caption{(Color online) Deformation parameters $\beta_2$ of Ne
nuclei as a function of the neutron number in comparison with
experimental results~\cite{Ram01}. The calculated values
corresponding to the Skyrme forces T24, T44, T62, T66~\cite{Les07}
effective interactions. The calculated data employing the
AMD~\cite{Min12,Sum12} are also shown for comparison.}\label{fig1}
\end{figure}

Fig.~1 illustrates the evolution of the axially-symmetric
deformations with neutron number along the neon isotopic chain for
various effective interactions, in comparison with experimental
results~\cite{Ram01}. The deformation parameter is defined in the
cylindrical coordinate as the expectation values of radius operators
$\beta_2 = \sqrt{\pi\over5} {\langle 2z^2-r^2 \rangle\over \langle
z^2+r^2 \rangle}$ where the optimal ones are found by minimizing the
total energy of the nucleus. If $\beta_2$ $>$ 0 that means the
nucleus with a prolate shape while $\beta_2$ $<$ 0 means an oblate
one. Our selected parameterizations include: The Skyrme force T24
with a substantial like-particle coupling constant $\alpha$ and a
vanishing proton-neutron coupling constant $\beta$; T62 with a large
proton-neutron coupling constant $\beta$ and a vanishing
like-particle coupling constant $\alpha$; and T66 with large and
equal proton-neutron and like-particle tensor-term coupling
constants. The calculated results of T44, which has a mixture of
like-particle and proton-neutron tensor terms, are found to be
practically the same with those of T62 and not shown here in
Fig.~\ref{fig1}. The calculated results employing the
AMD~\cite{Min12,Sum12} with a Gogny-D1S interaction are also shown
for comparison. Notice that the AMD calculations do not take into
account the pairing interaction.

In general, among all the isotopes, the deformations of $^{20}$Ne
and $^{24}$Ne depend much on the interactions adopted. The softness
of these nuclei can be understood because they have shallow energy
surfaces (shown in Ref.~\cite{Sag04} with a different Skyrme
interaction SGII), and results are easily changed by a delicate
balance among the contributions of energy density functionals in
different parameterizations. The sensitivity of $^{20}$Ne and
$^{24}$Ne was found also by the relativistic Hartree-Fock-Bogoliubov
(RHFB) model~\cite{Ebr11}. A decisive theoretical ingredient is the
fine structure of the s.p. spectra of those nuclei. Our interest and
focus in this work are to understand how spherical shapes of
$^{26}$Ne and $^{30}$Ne are predicted for all of the various
parameterizations shown in Fig.~\ref{fig1} (The $^{28}$Ne nucleus
might be triaxially-deformed~\cite{Min12,Sum12} which is beyond the
scope of this work). As mentioned in the introduction, those results
are not consistent with the large deformations found in
experiments~\cite{Ram01,Ril03,Yan03,Gei05,Iwa05,Gib07}, and the
inconsistency is also present in other theoretical mean-field models
such as the Hartree-Fock-Bogoliubov (HFB) theory~\cite{Hil07} and
the RHFB theory~\cite{Ebr11}. A better agreement of the AMD results,
which include no pairing, suggests the importance of not only
pairing correlations on the evolution of deformation, but also extra
contributions beyond mean-field models such as the
particle-vibration couplings, which needs to be studied in the
future. Hereafter, within the present model, we demonstrate how
sensitively the shapes of these nuclei depend on the pairing
strength together with the cooperative tensor correlation.

\begin{figure}[t!]
\centering
\includegraphics[width=13.5cm]{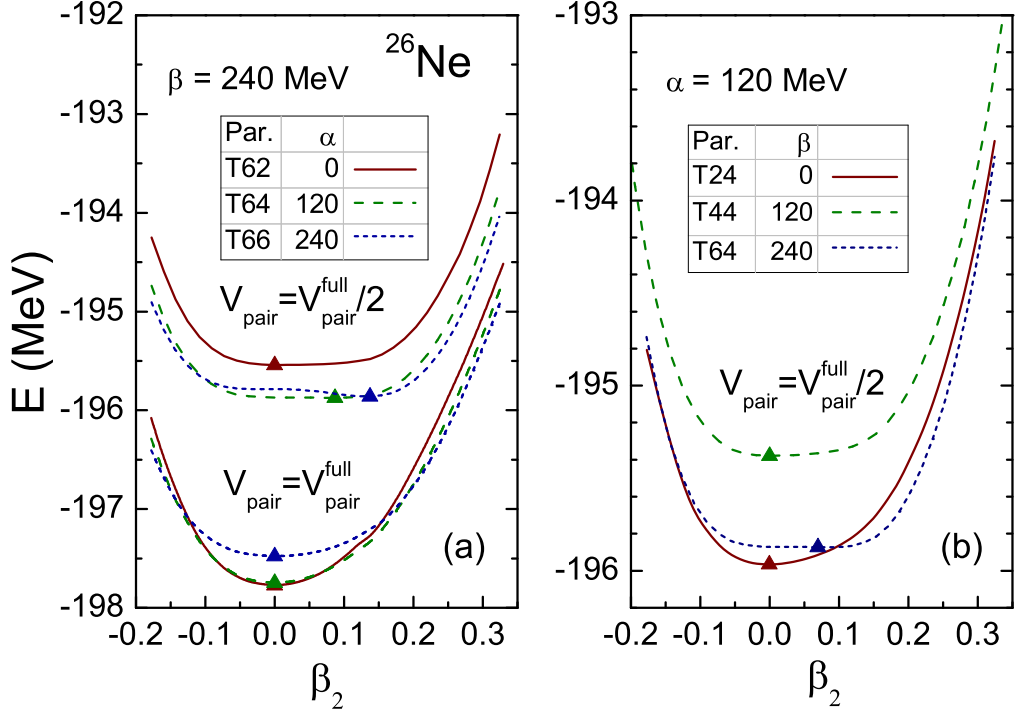}
\caption{(Color online) Energy surfaces of $^{26}$Ne as a function
of the quadruple deformation parameter $\beta_2$ using T62 (solid
line), T64 (dashed line), T66 (dotted line). The upper three curves
are done using a strong full pairing of $V_{\text{pair}}$ =
$V_{\text{pair}}^{\rm full}$, and the lower three curves are with a
medium pairing of $V_{\text{pair}}$ = $V_{\text{pair}}^{\rm
full}/2$. The energy minima are indicated with
triangles.}\label{fig2}
\end{figure}

In Fig.~\ref{fig2}, the energy surfaces of $^{26}$Ne are shown as a
function of the deformation parameter $\beta_2$ with increasing
tensor couplings: left panel for increasing values of like-particle
coupling $\alpha$ with T62 (solid line), T64 (dashed line), T66
(dotted line) at fixed proton-neutron coupling of $\beta = 240 $
MeV, and right panel for increasing proton-neutron coupling $\beta$
with T24 (solid line), T44 (dashed line), T64 (dotted line) at fixed
like-particle coupling of $\alpha = 120$ MeV. The calculations are
done using both a strong full pairing of $V_{\text{pair}}$ =
$V_{\text{pair}}^{\rm full}$, and a medium pairing of
$V_{\text{pair}}$ = $V_{\text{pair}}^{\rm full}/2$. The neutron
pairing gaps in the latter cases are only a few keV, and are
regarded as the cases in which we do one variant of calculations
with a weakened pairing to study its influence on the nuclear
deformation. The energy minima are indicated with triangles. Three
cases with $V_{\text{pair}}$ = $V_{\text{pair}}^{\rm full}$ all give
a spherical shape for $^{26}$Ne, even with increasingly strong
tensor forces. While in the case of $V_{\text{pair}}$ =
$V_{\text{pair}}^{\rm full}/2$, small $\alpha$ value in the case of
T62 gives a spherical minima, but larger $\alpha$ values in cases of
T64 and T66 drive clear deformations in the prolate side, which
indicates an essential role of $\alpha$ tensor strength on the shape
of $^{26}$Ne. Similarly, a larger $\beta$ value results in a prolate
deformation, as seen in the right panel of Fig.~\ref{fig2}. For
example, $\beta_2 = 0.087$ for T64 with a large value of $\beta =
240$ MeV with $\alpha = 120$ MeV. Therefore, to obtain an
experimentally observed prolate shape of $^{26}$Ne, a relatively
large tensor strength is obviously necessary, together with a
weakened pairing between nucleons. We mention here that this prolate
result was predicted before only by theoretical calculations based
on the RHFB model using one parameter set of PKO3 with $\beta_2 \sim
$ 0.2~\cite{Ebr11}.

\begin{figure}[t!]
\centering
\includegraphics[width=13.5cm]{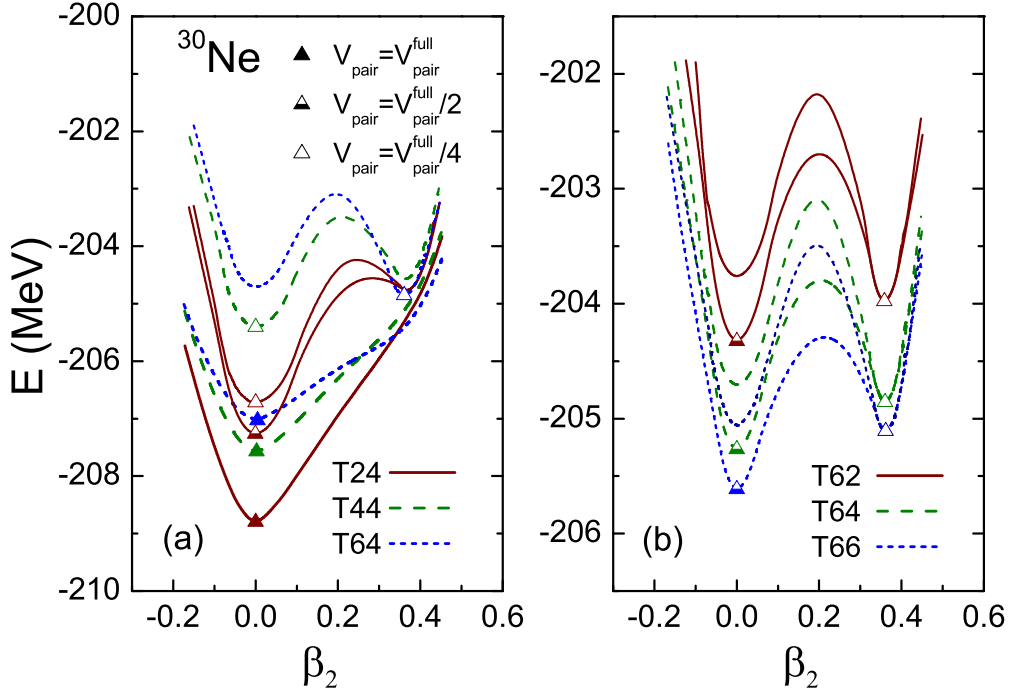}
\caption{(Color online) Same with Fig.~\ref{fig2} but for $^{30}$Ne.
The energy minima are indicated with triangles, and three cases of
pairing strengths are employed: a strong pairing case with
$V_{\text{pair}}$ = $V_{\text{pair}}^{\rm full}$ (filled symbols) ,
a medium one with $V_{\text{pair}}$ = $V_{\text{pair}}^{\rm
full}$~/2 (half symbols) and a weak one with
$V_{\text{pair}}$=$V_{\text{pair}}^{\rm full}$~/4 (open symbols). In
the left panel, all three pairing cases are shown for T24, and two
pairing cases ($V_{\text{pair}}$ = $V_{\text{pair}}^{\rm full}$,
$V_{\text{pair}}^{\rm full}$~/4) are shown for T44 and T64; In the
right panel, two pairing cases ($V_{\text{pair}}$ =
$V_{\text{pair}}^{\rm full}$~/2, $V_{\text{pair}}^{\rm full}$~/4)
are shown for T62, T64 and T66.}\label{fig3}
\end{figure}

For the case of $^{30}$Ne, we prepare in Fig.~\ref{fig3} its energy
surfaces using the same parameter sets in Fig.~\ref{fig2}. The
energy minima are indicated with triangles, and three cases of
pairing strengths are employed: a strong pairing case with
$V_{\text{pair}}$ = $V_{\text{pair}}^{\rm full}$ (filled symbols), a
medium one with $V_{\text{pair}}$ = $V_{\text{pair}}^{\rm full}$~/2
(half symbols) and a weak one with
$V_{\text{pair}}$=$V_{\text{pair}}^{\rm full}$~/4 (open symbols).
From the comparison of the T24 results with those of T44 and T64, we
see that a stronger tensor coupling in the self-consistent SHF
calculation makes the energy surface shallower, and at the same time
makes the second prolate minimum more pronounced. However, the
spherical minimums still win in the case of $V_{\text{pair}}$ =
$V_{\text{pair}}^{\rm full}$ (filled symbols), and in the case of
$V_{\text{pair}}$ = $V_{\text{pair}}^{\rm full}$~/2 (half symbols)
as well, even for the latter case, a weakened pairing helps to lift
largely in energy the spherical minimums. Interesting results are
obtained with a further weakened pairing in the case of
$V_{\text{pair}}$ = $V_{\text{pair}}^{\rm full}$~/4 (open symbols).
In this circumstance, T24 and T44 with smaller tensor strengthes
give still no deformed minima, but T62, T64, T66 with larger tensor
couplings finally achieve a large prolate-deformed shape at $\beta_2
\sim 0.35$ for $^{30}$Ne, as desired by the experiments. This
suggests that it demands the cooperation of a small pairing strength
and a large tensor force to obtain a large prolate deformation for
$^{30}$Ne in this SHF + BCS model. This is consistent with the
conclusion drawn in the context of $^{26}$Ne.

Our results demonstrate very clearly that the nucleon pairing
together with the tensor correlation are responsible to evolve the
shape of nuclei. The decisive role of a weak nucleon pairing can be
understood from a well-known fact that the pairing interaction tends
to form the $J = 0^+$ pairs of identical particles which have
spherically symmetric wave functions. The appearance of
well-deformed local minima in the weak pairing cases may be
indicated in the corresponding s.p. configurations at the minima. To
take T64, as an example, $[330~1/2]$ level from $1f_{7/2}$ orbit is
occupied with the occupation probability $v^2= 1.0$, while
$[202~3/2]$ level from $1d_{3/2}$ orbit is unoccupied with $v^2 =
0.0$. Here one-particle levels are given in the standard notation of
asymptotic quantum numbers $[Nn_z\Lambda\Omega]$. Those
configurations largely differ in the strong pairing case, where the
corresponding occupation probability for $1f_{7/2}$ orbit is $0.05$,
and that for $1d_{3/2}$ orbit is $0.90$. Meanwhile, the importance
of tensor interaction for generating deformed minima is due to the
fact that the tensor interaction brings in general reduced neutron
and proton shell gaps and enhanced s.p. level densities near the
Fermi level. To take T24 and T64 as a comparison, the
$1f_{7/2}-1d_{3/2}$ neutron gap is reduced from 4.49 MeV to 3.76
MeV, and the $1d_{5/2}-1p_{1/2}$ proton gap is reduced from 8.82 MeV
to 7.73 MeV.
\begin{figure}[t!]
\centering
\includegraphics[width=13.5cm]{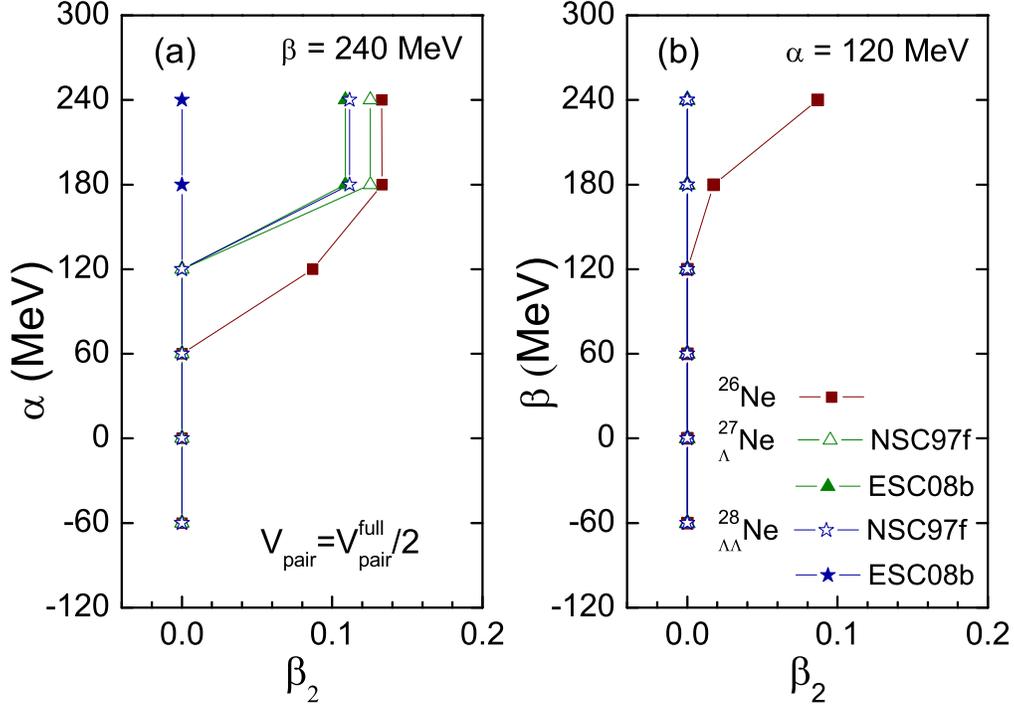}
\caption{(Color online) Left panel: Deformation changing of
$^{26}$Ne with increasing tensor coupling strength $\alpha$ using
T61, T62, T63, T64, T65, T66, for $\beta = 240 $ MeV; Right panel:
Deformation changing of $^{25}$Ne with increasing tensor coupling
strength $\beta$ using T14, T24, T34, T44, T54, T64, for $\alpha =
120 $ MeV. The calculations are done with a medium pairing of
$V_{\text{pair}}$ = $V_{\text{pair}}^{\rm full}/2$. The
corresponding singe- and double-hypernuclei are also presented in
dashed and dotted lines with two hyperon interactions: the ESC08b
model (filled symbols) and the NSC97f model (open
symbols).}\label{fig4}
\end{figure}

\begin{figure}[t!]
\centering
\includegraphics[width=13.5cm]{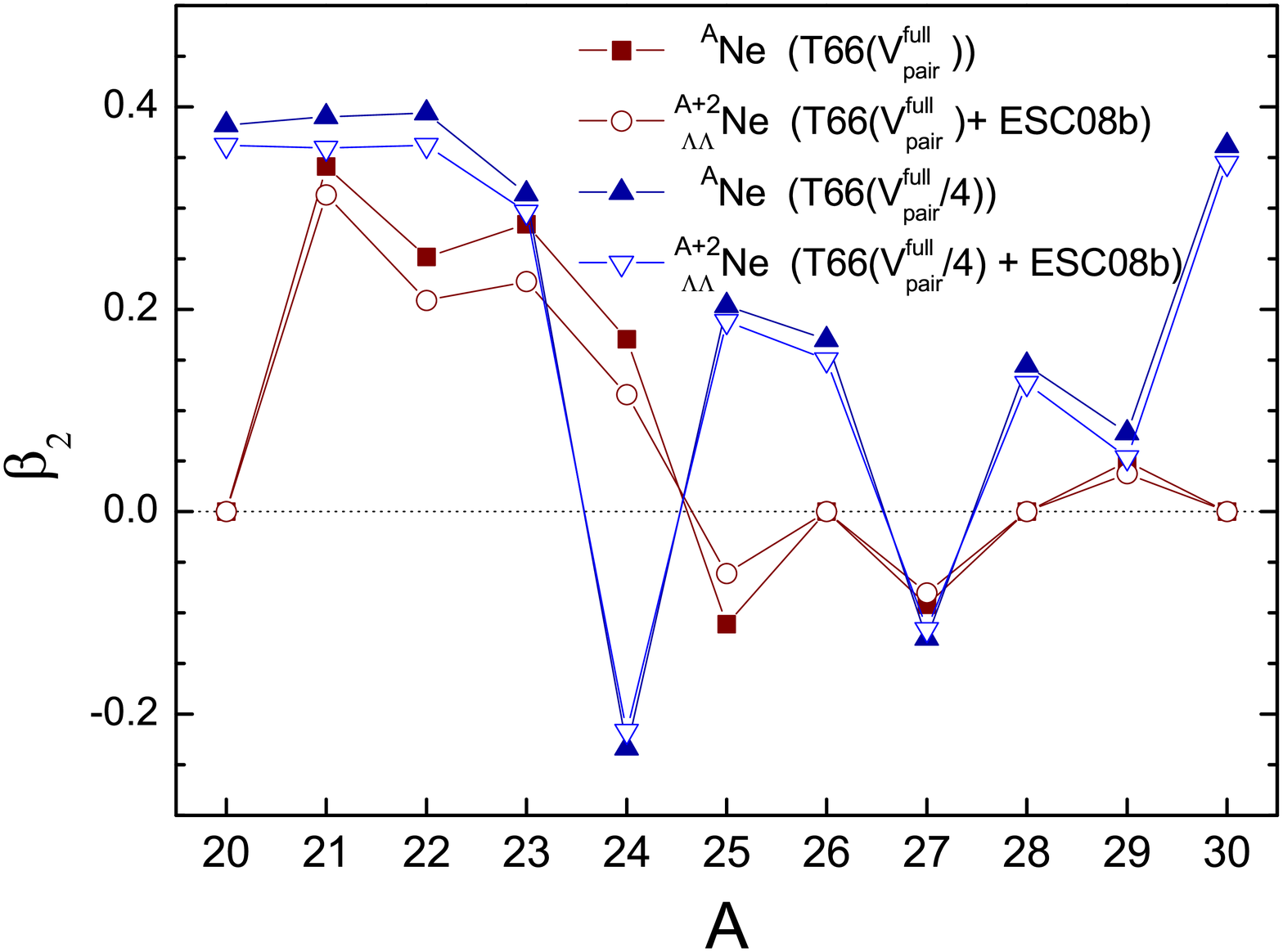}
\caption{(Color online) Deformation parameters of the double-lambda
hypernuclei (filled symbols) with neon isotopic core $^{20-30}$Ne
are plotted using the ESC08b potential, with the comparison of those
of core nuclei (open symbols). The calculations are done with T66
for two cases of pairing strengths: a full pairing of
$V_{\text{pair}}$ = $V_{\text{pair}}^{\rm full}$, and a weak pairing
of $V_{\text{pair}}$ = $V_{\text{pair}}^{\rm full}/4$.}\label{fig5}
\end{figure}

In order to illustrate more clearly the shape change due to the
tensor force in $^{26}$Ne, we present in Fig.~\ref{fig4} a
correlation between tensor coupling strength $\alpha$ or $\beta$ and
deformation $\beta_2$ at the energy minimum of $^{26}$Ne at fixed
$\beta = 240$ MeV (left panel) and $\alpha = 120$ MeV (right panel),
in the case of a medium pairing with $V_{\text{pair}}$ =
$V_{\text{pair}}^{\rm full}/2$. The corresponding singe- and
double-hypernuclei are also presented with two hyperon interactions:
the ESC08b model (filled symbols) and the NSC97f model (open
symbols). In general, the addition of the $\Lambda$ particle results
in a slightly smaller deformation, as is the same with the earlier
study~\cite{Zho07} with the absence of the tensor force. And from
Fig.~\ref{fig4}(a) we notice that with the ESC08b model hypernuclei
tend to be more spherical, which has its root in a deeper $\Lambda$
potential for ESC08b than for NSC97f as mentioned before. Specially,
due to the weak deformation minimum in this medium pairing case, the
double-$\Lambda$ hypernuclei $^{28}_{\Lambda\Lambda}$Ne is found to
be spherical when including the contribution of strong YN
interaction of ESC08b, compared with a prolate $^{26}$Ne core.

Our final results are presented in Fig.~\ref{fig5}, where the
deformation parameters of the double-lambda hypernuclei (open
symbols) with neon isotopic core $^{20-30}$Ne are plotted using the
ESC08b potential, with the comparison of the data of those of core
nuclei (filled symbols). The calculations are done with T66 for two
cases of pairing strengths: a full pairing of $V_{\text{pair}}$ =
$V_{\text{pair}}^{\rm full}$, and a weak pairing of
$V_{\text{pair}}$ = $V_{\text{pair}}^{\rm full}/4$. The softness of
$^{20,24}$Ne is again present, and we also find an interesting shape
inverse of $^{25}$Ne from oblate to prolate with the modification of
pairing strength. And prolately-deformed ground states are
successfully realized for $^{26,28,30}$Ne as a combined effect of a
large tensor force and a weakened pairing in the present model. This
fact might be quite meaningful for further improvements of the SHF
model or Skyrme parameterizations toward a better description on the
shell structures of nuclei in general. As was stated before, we see
that for all the isotopes there are smaller deformations with the
same shapes for hypernuclei, compared with corresponding core nuclei
in both the full pairing and the weak pairing cases.

\section{Summary and future perspectives}
In summary, we have performed  the deformed SHF +BCS model
calculations to investigate the effects of tensor and pairing forces
on the quadruple deformation of neon isotopes and the corresponding
$\Lambda$ hypernuclei. With selected parameterizations of various
tensor, pairing and hyperon-nucleon interactions, we disentangle the
interplay of these correlations for  the deformation of neon
isotopes and the corresponding hypernuclei. To investigate the role
of the pairing correlations, we adopt three kinds of the pairing
strength. For tensor interactions, we take 11 different Skyrme
parameter sets listed in Table 1. With these parameter sets, we
found in $^{25}$Ne and $^{26}$Ne the important interplay of the
tensor and pairing correlations which are different to those in the
lighter neon isotopes. Namely, in $^{21,22,23}$Ne, the ground states
are predicted always to be prolate deformed, irrespective to the
adopted interaction, which are consistent with both experiments and
previous calculations. The increase of tensor strength changes the
shape of $^{25}$Ne nucleus from oblate to prolate, and also the
shape of $^{26}$Ne nucleus from spherical to prolate with the help
of a weakened nucleon pairing interaction. The prolate shape of
$^{26}$Ne obtained with relatively large tensor strengths is quite
encouraging because it is consistent with the experimental $B(E2)$
data. We demonstrate also that the cooperation of a weakened pairing
and a large tensor interaction drives the shape of $^{30}$Ne from
spherical to prolate, as desired by the recent experiments.

In addition, the interplay of tensor force and hyperon force is also
studied, and the tensor effect on the deformation of the isotope is
found to be larger than that of $\Lambda$ particles added to the
core nucleus with realistic hyperon interactions. With the same core
nuclei, the ESC08b potential makes the corresponding single- and
double-hypernuclei harder to deform than the NSC97f model because of
a deeper $\Lambda$ potential depth of the former case.

As a future project, it is quite important to examine  further the
strengths of the realistic tensor interaction for the mean field
models, for which we may have to refer to more microscopic
calculations of the nuclear energy density functional based on
realistic NN interactions, such as chiral NN potential
N3LOW~\cite{Hol11,Kai12}. With that our calculation might be
 improved using a microscopic-determined tensor strength
parameters (which might be density-dependent) with more insight into
the tensor role on nuclear structures. Also, to clarify the pairing
effect, as mentioned in the introduction, we should update our
calculation with a recently proposed isospin-dependent pairing
force, using a  proper pairing strength parameters fitted from the
systematic experimental pairing gaps~\cite{Mar07,Yama12}. At the
same time, it might be important to accommodate the effects beyond
the mean-field model such as the particle-vibration coupling effect
in the future study.

\begin{acknowledgments}
We would like to thank Prof. J. Meng, Prof. K. Hagino and Dr. H.Z.
Liang for valuable discussions. This work is supported by a
Grants-in-Aid for Scientific Research from Monbukagakusho of Japan
(21540288 and 20105003), and by the National Natural Science
Foundation of China (Grant No 10905048, 10975116 and 11275160).

\end{acknowledgments}


\end{document}